%% file: acm-draft.tex
\documentclass[sigconf]{acmart}
\AtBeginDocument{%
  }

\usepackage{xspace}
\usepackage{enumitem}
\usepackage{algorithm}
\usepackage{algorithmic}
\usepackage{enumitem}
\usepackage{fancyvrb}
\usepackage{listings}
\usepackage{xcolor}
\usepackage{threeparttable}
\usepackage{pgfplots} % For bar chart
\usepackage{subcaption}
\newcommand{\ourtool}{\mbox{\textsc{DrillAgent}}\xspace}
\newcommand{\VAA}{\mbox{\textsc{VAAgent}}\xspace}
\newcommand{\ITA}{\mbox{\textsc{ITAgent}}\xspace}
\newcommand{\PEA}{\mbox{\textsc{PEAgent}}\xspace}
\newcommand{\CTA}{\mbox{\textsc{CTAgent}}\xspace}

\newcommand{\parabf}[1]{\noindent\textbf{#1}}
\newcommand{\CodeIn}[1]{{\small \texttt{#1}}}
\newcommand{\FormulaIn}[1]{{\small \texttt{$#1$}}}
\newcommand{\Comment}[1]{}

\setcopyright{none}
\begin{document}

%%
%% The "title" command has an optional parameter,
%% allowing the author to define a "short title" to be used in page headers.
% \title{DrillAgent: Towards Automated Proof-of-Vulnerability Generation via Execution-State-Aware LLM Reasoning}
\title{Execution-State-Aware LLM Reasoning for Automated Proof-of-Vulnerability Generation} % For arxiv

%%
%% The "author" command and its associated commands are used to define
%% the authors and their affiliations.
%% Of note is the shared affiliation of the first two authors, and the
%% "authornote" and "authornotemark" commands
%% used to denote shared contribution to the research.
\author{Haoyu Li}
\affiliation{%
  \institution{University of Illinois Urbana-Champaign}
  \city{Champaign}
  \state{Illinois}
  \country{USA}}
\email{haoyuli9@illinois.edu}

\author{Xijia Che}
\affiliation{%
  \institution{University of Illinois Urbana-Champaign}
  \city{Champaign}
  \state{Illinois}
  \country{USA}}
\email{xijiac2@illinois.edu}

\author{Yanhao Wang}
\affiliation{%
  \institution{Independent Researcher}
  \city{Shanghai}
  \state{Shanghai}
  \country{China}}
\email{wangyanhao136@gmail.com}

\author{Xiaojing Liao}
\affiliation{%
  \institution{University of Illinois Urbana-Champaign}
  \city{Champaign}
  \state{Illinois}
  \country{USA}}
\email{xjliao@illinois.edu}

\author{Luyi Xing}
\affiliation{%
  \institution{University of Illinois Urbana-Champaign}
  \city{Champaign}
  \state{Illinois}
  \country{USA}}
\email{lxing2@illinois.edu}

% \author{Ben Trovato}
% \authornote{Both authors contributed equally to this research.}
% \email{trovato@corporation.com}
% \orcid{1234-5678-9012}
% \author{G.K.M. Tobin}
% \authornotemark[1]
% \email{webmaster@marysville-ohio.com}
% \affiliation{%
%   \institution{Institute for Clarity in Documentation}
%   \city{Dublin}
%   \state{Ohio}
%   \country{USA}
% }

%%
%% By default, the full list of authors will be used in the page
%% headers. Often, this list is too long, and will overlap
%% other information printed in the page headers. This command allows
%% the author to define a more concise list
%% of authors' names for this purpose.
% \renewcommand{\shortauthors}{Trovato et al.} % Mute by LHY

%%
%% The abstract is a short summary of the work to be presented in the
%% article.
\input{sections/abstract}

\maketitle

% Main body of the paper.
\input{sections/1_introduction}

\input{sections/2_background}
\input{sections/3_design}
\input{sections/4_implementation}
\input{sections/5_evaluation}
\input{sections/6_discussion}
\input{sections/7_related_work}
\input{sections/8_conclusion}

%%
%% The acknowledgments section is defined using the "acks" environment
%% (and NOT an unnumbered section). This ensures the proper
%% identification of the section in the article metadata, and the
%% consistent spelling of the heading.
% Omitted for arxiv
% \begin{acks}
% To Robert, for the bagels and explaining CMYK and color spaces.
% \end{acks}

%%
%% The next two lines define the bibliography style to be used, and
%% the bibliography file.
\bibliographystyle{ACM-Reference-Format}
% \bibliography{sample-base}
\bibliography{acm-cites}

%%
%% If your work has an appendix, this is the place to put it.
% \appendix
% \section{Open Science}
% Required by CCS 2026.

% \section{Research Methods}

% \subsection{Part One}

% Lorem ipsum dolor sit amet, consectetur adipiscing elit. Morbi
% malesuada, quam in pulvinar varius, metus nunc fermentum urna, id
% sollicitudin purus odio sit amet enim. Aliquam ullamcorper eu ipsum
% vel mollis. Curabitur quis dictum nisl. Phasellus vel semper risus, et
% lacinia dolor. Integer ultricies commodo sem nec semper.

% \subsection{Part Two}

% Etiam commodo feugiat nisl pulvinar pellentesque. Etiam auctor sodales
% ligula, non varius nibh pulvinar semper. Suspendisse nec lectus non
% ipsum convallis congue hendrerit vitae sapien. Donec at laoreet
% eros. Vivamus non purus placerat, scelerisque diam eu, cursus
% ante. Etiam aliquam tortor auctor efficitur mattis.

% \section{Online Resources}

% Nam id fermentum dui. Suspendisse sagittis tortor a nulla mollis, in
% pulvinar ex pretium. Sed interdum orci quis metus euismod, et sagittis
% enim maximus. Vestibulum gravida massa ut felis suscipit
% congue. Quisque mattis elit a risus ultrices commodo venenatis eget
% dui. Etiam sagittis eleifend elementum.

% Nam interdum magna at lectus dignissim, ac dignissim lorem
% rhoncus. Maecenas eu arcu ac neque placerat aliquam. Nunc pulvinar
% massa et mattis lacinia.

\end{document}

%% file: sections/abstract.tex
\begin{abstract}
Proof-of-Vulnerability (PoV) generation is a critical task in software security, serving as a cornerstone for vulnerability validation, false positive reduction, and patch verification.
While directed fuzzing effectively drives path exploration, satisfying complex semantic constraints remains a persistent bottleneck in automated exploit generation.
Large Language Models (LLMs) offer a promising alternative with their semantic reasoning capabilities; however, existing LLM-based approaches lack sufficient grounding in concrete execution behavior, limiting their ability to generate precise PoVs.

In this paper, we present \ourtool, an agentic framework that reformulates PoV generation as an iterative \textit{hypothesis--verification--refinement} process.
To bridge the gap between static reasoning and dynamic execution, \ourtool synergizes LLM-based semantic inference with feedback from concrete program states.
The agent analyzes the target code to hypothesize inputs, observes execution behavior, and employs a novel mechanism to translate low-level execution traces into source-level constraints.
This closed-loop design enables the agent to incrementally align its input generation with the precise requirements of the vulnerability.
We evaluate \ourtool on SEC-bench, a large-scale benchmark of real-world C/C++ vulnerabilities.
Experimental results show that \ourtool substantially outperforms state-of-the-art LLM agent baselines under fixed budget constraints, solving up to 52.8\% more CVE tasks than the best-performing baseline.
These results highlight the necessity of execution-state-aware reasoning for reliable PoV generation in complex software systems.
\end{abstract}

%% file: sections/1_introduction.tex
\section{Introduction}
\label{sec:intro}

% Paragraphs:
% 1. Brief task definition. Usage scenarios and meanings (and different settings).
% 2. Existing traditional approaches. Research gap and limitations. 
% 3. LLM agent-solutions vs. fuzzing solutions. (tricky)
% 3. Main challenges for LLM agents.
% 4. Our approach.
% 5. Our contributions.

Vulnerability reproduction is a cornerstone of software security.
Given a reported vulnerability, a \textit{Proof-of-Vulnerability} (PoV) is a concrete input that deterministically triggers the vulnerability by violating an intended security property at a specific target site.
Reliable PoV generation is critical not only for filtering false alarms from static analysis~\cite{bao2025alarms} but also for serving as ground truth in downstream tasks such as patch validation in automated program repair (APR)~\cite{cheng2025brtagent, ahmed2025otter}.
However, despite decades of research, automating PoV generation for real-world software remains a persistent challenge, particularly for bugs buried deep within complex logic.

State-of-the-art approaches predominantly rely on \textit{directed greybox fuzzing} (DGF)\cite{bohme2017directed, chen2018hawkeye, huang2022beacon}.
By defining a distance metric between program states and a vulnerability site, DGF heuristically steers execution toward the target.
While effective for shallow vulnerabilities, prior work has shown that DGF struggles to reproduce deep, logic-heavy bugs guarded by complex semantic constraints\cite{bao2025alarms, feng2025randluzz}.
The fundamental limitation is that geometric distance is feasibility-unaware\cite{bai2025aflgopher}: a path may be syntactically short yet logically blocked by intricate sanitization checks or data dependencies (e.g., magic bytes, checksums, or protocol invariants).
As a result, fuzzers without semantic understanding often expend substantial effort mutating inputs against constraints that cannot be satisfied through coverage-oriented exploration alone.

Large Language Models (LLMs) theoretically offer the semantic reasoning capabilities needed to bridge this gap.
An LLM can infer input formats, reconstruct protocol logic, and hypothesize constraint-satisfying inputs directly from source code.
However, recent studies demonstrate that off-the-shelf LLMs perform poorly on end-to-end PoV generation tasks~\cite{lee2025sec}.
We argue that this failure is not merely due to model capacity but stems from a mismatch in problem formulation.
PoV generation is not a static translation problem but an interactive debugging process that requires reasoning over execution outcomes.
LLMs operate primarily in a conceptual space derived from static text and lack grounding in the program’s concrete runtime behavior.
Without observing how memory layouts, variable states, or branch conditions evolve during execution, LLMs tend to hallucinate inputs that appear plausible in code but fail under strict runtime semantics.

Ultimately, progress in PoV generation hinges on the ability to interpret execution outcomes semantically.
Bridging the gap between semantic hypothesis and execution reality requires addressing three key challenges:

% Direct CLI compare? No need to be use ablation study. 
\noindent\textbf{Challenge 1: Contextual Dependency Retrieval.}
Vulnerability triggers often depend on non-local context, such as state transitions defined across distant functions or header files.
Extracting the minimal yet sufficient code context governing a target constraint without overwhelming the LLM’s context window is non-trivial.

\noindent\textbf{Challenge 2: Semantic Gap in Execution Feedback.}
When a generated input fails, feedback from testing tools is typically low-level (e.g., coverage bitmaps, exit codes, or crash signals).
There is a semantic impedance mismatch: translating these opaque binary signals into high-level source constraints that an LLM can understand remains a major bottleneck.
% TODO: Refine here to make it understandable.
% In the design subsection, illustrate more about the gap.

% Precise State Satisfaction.
% Satisfaction of Vulnerability Triggering Constraints
\noindent\textbf{Challenge 3: Satisfaction of Vulnerability Constraints.} 
Reaching the target site is necessary but insufficient for a vulnerability proof.
Triggering vulnerabilities often requires satisfying precise and intricate runtime predicates—such as specific heap layouts or global state configurations—that are neither explicit in the local source code nor inferable without execution-aware reasoning.

\vspace{5pt}
\emph{Our Approach.}
To this end, we propose \ourtool, an agentic framework that automates PoV generation by tightly closing the loop between semantic hypothesis and execution-grounded verification.
Unlike prior approaches that treat LLMs as one-shot generators, \ourtool models PoV generation as an iterative \textit{hypothesis--verification--refinement} process.
The agent first analyzes the program to hypothesize a candidate input.
It then executes the program and employs a novel \textit{Trace-to-Prompt} translator to convert execution traces—specifically failed branch conditions mapped back to source code—into natural language constraints.
This execution-grounded feedback enables the agent to debug and refine its own hypotheses, incrementally converging on inputs that satisfy both reachability and vulnerability-triggering conditions.

In summary, this paper makes the following contributions:
\begin{itemize}[leftmargin=*, noitemsep, topsep=2pt]
    \item We present \ourtool, an end-to-end LLM agentic framework that automates Proof-of-Vulnerability generation by tightly coupling semantic reasoning with execution-state feedback. It models PoV generation as an iterative, closed-loop process, enabling the agent to reason about both reachability and vulnerability-triggering conditions in real-world C/C++ programs.
    % \textbf{End-to-end Agentic Framework.} We present \ourtool, the first framework to tightly integrate LLM-based semantic reasoning with directed execution feedback for automated PoV generation.
    \item We introduce an LLM-agent-friendly execution feedback mechanism that bridges the semantic gap between high-level LLM reasoning and low-level program execution. Our design integrates automated code instrumentation, fine-grained coverage feedback, and sanitizer-based crash validation into a unified feedback loop. Execution outcomes are translated into source-level, semantically interpretable constraints, allowing LLMs to iteratively debug failed hypotheses and align their reasoning with concrete runtime behavior.
    % \textbf{Execution Feedback Mechanism.} We introduce a technique that translates low-level execution traces into semantic prompts, enabling LLMs to learn effectively from runtime failures.
    \item We evaluate \ourtool on SEC-bench, a large-scale benchmark of real-world C/C++ vulnerabilities. Experimental results show that \ourtool outperforms state-of-the-art LLM agent baselines under fixed budget constraints, solving up to 52.8\% more CVE tasks than the previously best-performing baseline, and successfully reproducing a substantial number of previously unresolved benchmark vulnerabilities.
    % \textbf{Evaluation.} We evaluate \ourtool on real-world CVEs and show that it outperforms state-of-the-art LLM agent baselines, improving PoV reproduction success rates by up to \revision{XX\%}.
\end{itemize}

%% file: sections/2_background.tex
\section{Background and Motivation}

\subsection{Problem Statement}

% The goal of automated Proof-of-Vulnerability (PoV) generation is to synthesize a concrete input that reliably demonstrates the exploitability of a reported vulnerability.
% Formally, let $P$ denote the target program and $\mathcal{I}$ be the universe of all possible inputs.
% A vulnerability is modeled as a tuple $\mathcal{V} = \langle L_{target}, \Phi_{crash} \rangle$, where $L_{target}$ identifies the vulnerable program location (e.g., a basic block or source line), and $\Phi_{crash}$ represents the failure condition (e.g., memory safety violation or assertion failure).

% Given a test harness $H$ (which may include sanitizers), the execution of $P$ on input $I \in \mathcal{I}$ is denoted as $Exec(P, H, I)$.
% The objective is to find an input $I^* \in \mathcal{I}$ such that the execution trace $\tau = Exec(P, H, I^*)$ satisfies two conditions:
% \begin{itemize}
%     \item (1) Reachability: The trace $\tau$ covers $L_{target}$;
%     \item (2) Triggering: The program state at $L_{target}$ satisfies $\Phi_{crash}$.
% \end{itemize}

% \parabf{Research Scope.} In real-world C/C++ programs, finding such an $I^*$ is non-trivial because the subspace of valid inputs triggering the vulnerability is often extremely sparse and guarded by complex path constraints.
% In this work, we focus on memory-related vulnerabilities, as they represent a critical class of security threats with well-defined validation oracles (e.g., AddressSanitizer), though our formulation is generic to other bug types.

The objective of automated Proof-of-Vulnerability (PoV) generation is to synthesize a concrete input that demonstrates the existence and exploitability of a reported security flaw in a target program. 
For the program \FormulaIn{P}, a reported vulnerability is characterized by a tuple \FormulaIn{\mathcal{V} = \langle V_{location}, V_{effect} \rangle}, where:
\begin{itemize}[leftmargin=*]
  \item \FormulaIn{V_{location}} represents the target vulnerability location (e.g., a particular basic block or source line).
  \item \FormulaIn{V_{effect}} describes the expected effect of the vulnerability (e.g., a crash signal, data corruption, or information leakage).
\end{itemize}

Let $H$ be a test harness that encapsulates program $P$ and provides the necessary execution environment. 
Let $I \in \mathcal{I}$ be a candidate input.
We denote by $Exec(P, H, I)$ the execution of $P$ under harness $H$ with input $I$. 
Validation oracle $\mathcal{O}$ determines whether the execution triggers the target vulnerability as specified by $\mathcal{V}$.

\smallskip
\parabf{PoV Generation Task.} The goal is to find an input $I$ such that the execution $Exec(P, H, I)$ satisfies the following conditions, as verified by the oracle $\mathcal{O}$:

% \noindent
(1) {Target Reachability}: the execution must successfully reach the specified vulnerability location $V_{location}$.

% satisfy the necessary constraints to
(2) {Triggering Effect}: upon reaching the target location, the input must  trigger the vulnerability with an observable effect $V_{effect}$.
% Add a formula here.
Formally, the oracle evaluates
$$\mathcal{O}(P, \mathcal{V}, H, I) \rightarrow \{Success, Failure\}$$

% Our research scope.
\parabf{Research Scope.} In this work, we focus on PoV generation for real-world vulnerabilities in C/C++ open-source projects. 
Although current \ourtool mainly works on memory-related vulnerabilities, it can be extended to logical vulnerability scenarios.
We select this scope for two primary reasons. 
First, C/C++ programs are ubiquitously deployed in critical infrastructure systems, rendering the reproduction of their vulnerabilities highly impactful. 
Second, memory vulnerabilities in C/C++ leverage mature and robust validation oracles $\mathcal{O}$, particularly memory sanitizers that have undergone extensive practice over time.

%%%%%%%%%%%%%%%%%%%%%%%%%%%%%%%%%%%%%%%%%%%%%%%%%%%%%%%%%%%%%%%%%%%%%%%%%

\subsection{Motivating Example}
\label{sec:motivating-example}

% We use \texttt{CVE-2023-0760}, a heap-buffer-overflow in the popular multimedia framework GPAC~\cite{le2020gpac}, to illustrate the challenges of end-to-end PoV generation, and how \ourtool addresses the aforementioned three challenges. %raised in Section~\ref{sec:intro}. 

\begin{figure}[h]
  \centering
    
\begin{subfigure}[t]{0.95\linewidth}
\captionsetup{skip=2pt}
\includegraphics[width=\linewidth]{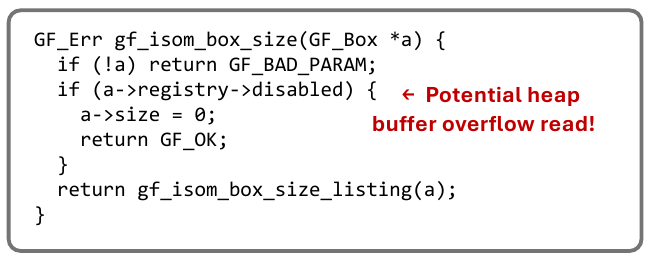}
\caption{Potential heap-buffer-overflow read.}
\label{fig:gpac-bof}
\end{subfigure}
\vspace{0.4em} 

\begin{subfigure}[t]{0.95\linewidth}
\captionsetup{skip=2pt}
\includegraphics[width=\linewidth]{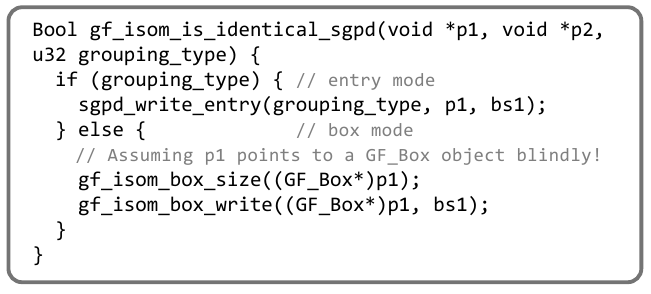}
\caption{Type confusion due to blind pointer casting.}
\label{fig:gpac-type-confusion}
\end{subfigure}

\caption{gpac.cve-2023-0760: vulnerable code snippets.}
\label{fig:motivating-example}
\end{figure}

We use CVE-2023-0760 in the GPAC multimedia framework as a motivating example to illustrate the difficulty of generating end-to-end PoVs. This vulnerability is triggered only after a long execution chain that includes MP4 container parsing, fragment-merging logic, and a comparison helper with unsafe type assumptions. 
The final crash happens in \CodeIn{gf\_isom\_box\_size} (Figure~\ref{fig:gpac-bof}), yet the true root cause lies earlier in the call trace: an object originating from the \CodeIn{sgpd} parsing stage is misinterpreted as a \CodeIn{GF\_Box} due to a blind pointer cast (Figure~\ref{fig:gpac-type-confusion}). 
This case therefore represents a typical multi-stage semantic bug rather than a local memory error.

\vspace{5pt}
\parabf{Vulnerability details.}
The type confusion occurs in the helper function shown in Figure~\ref{fig:gpac-type-confusion}. Function \CodeIn{gf\_isom\_is\_identical\_sgpd} accepts two opaque pointers of type \CodeIn{void*}. When the parameter \CodeIn{grouping\_type == 0} (the “box mode”), the code directly casts \CodeIn{p1} to \CodeIn{GF\_Box*} and invokes box-related routines without any validation. In practice, however, the \CodeIn{sgpd} parsing logic allocates sample group entry objects whose memory layout differs from that of a real \CodeIn{GF\_Box}. Passing such an group entry object to this branch violates the expected type invariant and creates a confused pointer.

Once this confused pointer reaches the function in Figure~\ref{fig:gpac-bof}, \CodeIn{gf\_isom\_box\_size} assumes that its argument is a \CodeIn{GF\_Box} and immediately dereferences nested fields such as \CodeIn{a->registry->disabled}. Because the underlying object is not a box instance, these offsets correspond to memory outside the allocated region of the entry object. The access therefore escapes the object boundary and results in a heap out-of-bounds read.

Triggering this bug requires several non-local conditions to be satisfied simultaneously. The input must be a well-formed fragmented MP4 file containing boxes such as \CodeIn{moof}, \CodeIn{traf}, and \CodeIn{sgpd} so that fragment merging is exercised. During merging, the comparison helper must be invoked with \CodeIn{grouping\_type == 0}, and the pointer must point to an \CodeIn{sgpd} entry rather than a genuine \CodeIn{GF\_Box}. Only under this specific combination does the type confusion propagate into \CodeIn{gf\_isom\_box\_size} and materialize as the memory error.

\vspace{5pt}
% \parabf{Why this breaks vanilla LLM agents.}
\parabf{Challenges for general-purpose LLM agents.}
This case highlights fundamental limitations of general-purpose agents: generating a valid PoV requires reasoning over constraints that span across the codebase and are not explicit in local code contexts.
% This CVE is representative because generating a valid PoV requires simultaneously satisfying multiple constraints, while the relevant logic is scattered across the codebase.

First, the agent must recover \emph{non-local contextual dependencies}.
Reaching the vulnerability requires combining MP4 container semantics (a fragmented file with \CodeIn{moof}, \CodeIn{traf}, and \CodeIn{sgpd} boxes), allocation behavior during \CodeIn{sgpd} parsing, and fragment-merge logic that eventually invokes the comparison helper.
Simple constraint-collection approaches often miss one of these steps, producing either invalid inputs rejected early or valid MP4 files that never exercise the vulnerable path.

Second, normal execution feedback provides little semantic guidance.
Common LLM agents typically can only observe program exit code and outputs, including ``invalid file,'' ``no fragments,'' or simply no crash.
Translating these superficial signals into actionable insights—such as identifying that fragment merging failed to trigger despite successful parsing—is almost impossible. This task requires more fine-grained execution state feedback to bridge a semantic gap that most general-purpose LLM agents are unable to navigate.

Third, reaching the vulnerable function is not sufficient.
To trigger the out-of-bounds read, execution must enter the ``box mode'' of the
comparison helper while passing a pointer that does not correspond to a real
\CodeIn{GF\_Box}.
At the same time, the input must be a fragmented MP4 so that the merge logic is
executed at all.
These precise runtime predicates are difficult to infer from local code alone,
making the final crash-triggering step particularly challenging.

% \smallskip
\vspace{5pt}
\parabf{How \ourtool addresses it.}
\ourtool successfully generated a validated PoV for this
vulnerability in our experiments.
During \emph{path exploration}, it learned the required input file structure
and produced well-formed fragmented MP4 files that reach the vicinity of the
vulnerable code.
During \emph{crash triggering}, \ourtool reused these near-target inputs and focused
on satisfying the remaining type-confusion condition, finally triggering the crash in
the third refinement round.

In contrast, an OpenHands + Claude Sonnet~4.5 baseline failed to generate a PoV after 74 iterations with unlimited budget.
Its attempts oscillated between malformed inputs that fail parsing and valid MP4 files that do not satisfy the fragment-merge constraints needed to reach the unsafe cast.
This behavior directly reflects the challenges above: missing non-local dependencies, limited use of execution feedback, and difficulty converging on the final vulnerability-triggering condition.

%% file: sections/3_design.tex
\section{Design of \ourtool}

% Exhibit our innovation points.
% 1. Make the execution transparent with fine-grained coverage feedback.
% 2. PoV generation is not simply constraints solving. Understand the vulnerability semantically and in the high-level is necessary.
% 3. It is non-trivial to teach LLM how to operate on binary files.
% 4. Path exploration and crash triggering should be separated.

\begin{figure*}[t]
  \centering
  \includegraphics[width=\textwidth]{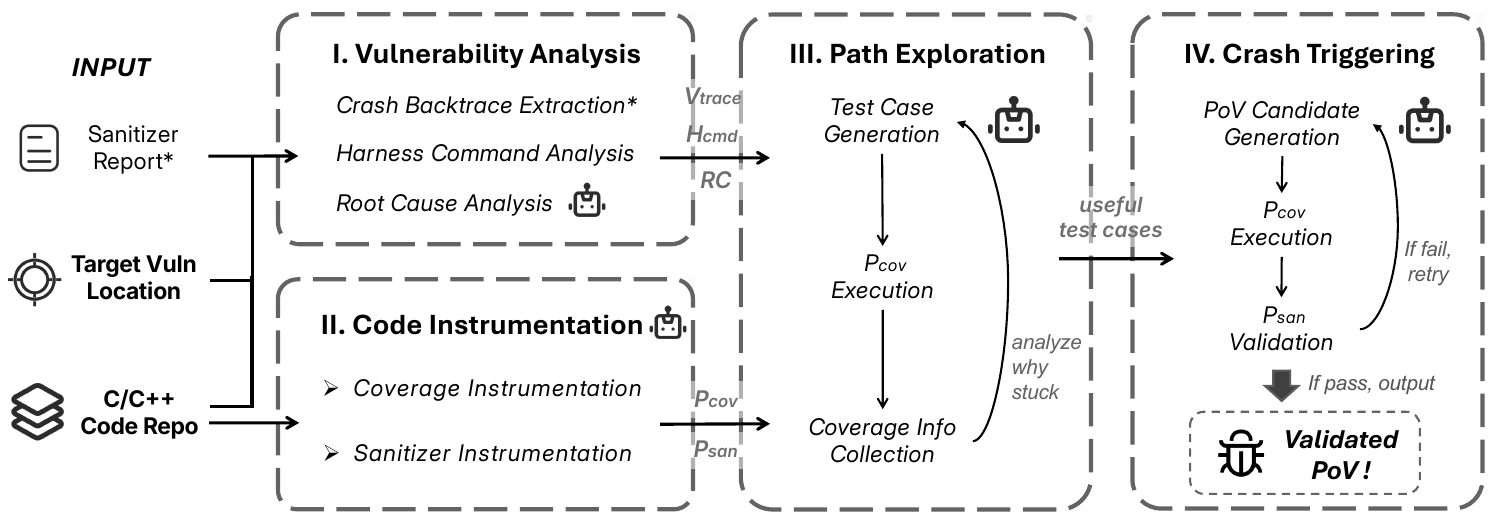}
  \caption{The Overall Workflow of \ourtool.}
  \Description{An overview of the \ourtool workflow, showing the agent-driven pipeline for PoV generation.}
  \label{fig:overview}
\end{figure*}

\subsection{Design Overview}
% \parabf{Overview.}
Figure \ref{fig:overview} shows the end-to-end workflow of \ourtool. 
It takes the source code repository and the target vulnerability location as mandatory inputs, with optional auxiliary vulnerability information provided (e.g., a sanitizer report). Through a fully automated LLM-driven reasoning and analysis process, \ourtool will ultimately output a validated PoV that can reliably trigger the vulnerability crash at the specified location.

\ourtool consists of four sequential phases. %, each implemented as a role-specialized sub-agent with a distinct objective and prompt configuration.
First, in the vulnerability analysis phase (\S~\ref{subsec:vuln_analysis}), \ourtool fully leverages the provided inputs to obtain a deep understanding of the given vulnerability, including its root cause (\FormulaIn{RC}) and the harness execution command (\FormulaIn{H_{cmd}}) required to trigger it. If a sanitizer report is available, \ourtool further extracts and refines the crash backtrace to derive the \FormulaIn{V_{trace}}.
Then, in the code instrumentation phase (\S~\ref{subsec:instrumentation}), two separate copies of the source code repository are compiled with coverage instrumentation and memory-sanitizer instrumentation respectively, yielding two corresponding binary programs (\FormulaIn{P_{cov}} and \FormulaIn{P_{san}}) for different subsequent purposes.
In the path exploration phase (\S~\ref{subsec:path_explore}), \ourtool iteratively attempts to generate test cases that satisfy the input format requirements, with the goal of reaching the vulnerable function at the branch level as much as possible. By executing \FormulaIn{H_{cmd}} on \FormulaIn{P_{cov}}, \ourtool can extract fine-grained coverage information from the current execution, which provides insights for the next iteration.
Finally, in the crash triggering phase (\S~\ref{subsec:crash_trigger}), based on test cases that have already reached or come very close to the crash location, \ourtool aggressively attempts to generate PoV candidates aiming to trigger the vulnerability crash, and executes \FormulaIn{H_{cmd}} on \FormulaIn{P_{san}} to validate whether they indeed violate memory safety. 
The PoV generation task is considered successful if the sanitizer reports a crash with a type and location consistent with the ground truth.

Each of the above four phases is realized by a role-specialized sub-agent with a distinct objective and prompt configuration, all coordinated by \ourtool within a unified execution workflow. For clarity, we refer to these sub-agents as the Vulnerability Analysis Agent (\VAA), the Code Instrumentation Agent (\ITA), the Path Exploration Agent (\PEA), and the Crash Triggering Agent (\CTA), respectively.

% Materials beyond text:
% 1. data format example of intermediate results
% 2. algorithm table
% 3. prompt snippet
% 4. unordered list enumeration

\subsection{Vulnerability Analysis}
\label{subsec:vuln_analysis}
% VAA主要做三件事：
% Crash Backtrace Extraction（即使有sanitizer report通常也很难直接用，因为编译预处理和优化导致行号不准）（多traces处理）
% Harness command analysis（文件后缀要求的问题）
% Root Cause Analysis（最重要！critical steps识别引导）
% 素材：Crash backtrace的JSON数据结构示例，等有了motivating example了再加。

In the first phase of the workflow, we use \VAA to analyze and interpret the target vulnerability based on all provided inputs, transforming heterogeneous raw information into multiple formatted key elements for PoV generation. The resulting vulnerability-related insights will guide the subsequent contextual dependency retrieval, enabling the agent to focus its attention effectively {(addressing Challenge 1)}. In particular, we focus on three critical facets that jointly constitute the semantic foundation for subsequent reasoning.

\begin{figure}
    \centering
    \captionsetup{skip=5pt}
    \includegraphics[width=0.9\linewidth]{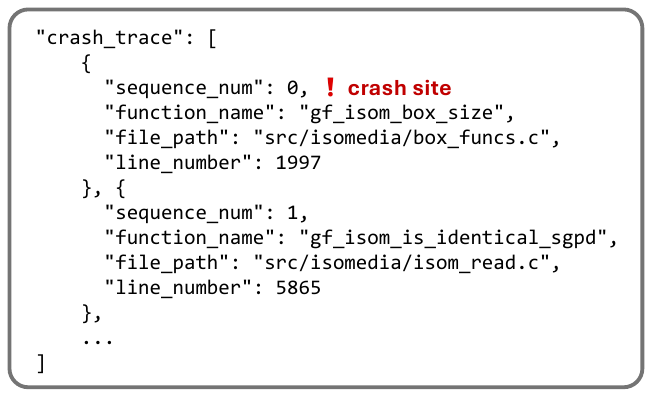}
    \caption{Simplified crash trace generated by the \VAA.}
    \label{fig:crash-trace}
    \Description{Simplified crash trace generated by the vulnerability analysis agent.}
\end{figure}

\smallskip
\parabf{Crash Backtrace Extraction.} Assuming the provided inputs include a sanitizer report of the target vulnerability, \VAA extracts the crash backtrace from the program entry location to the crash location and converts it into a JSON format (as shown in the Listing~\ref{fig:crash-trace}). While sanitizer reports typically contain such information, source line numbers are often different between the LLM agent’s view and the memory sanitizer’s view due to compiler preprocessing and code optimizations. To address this, \VAA refines the line number of each stack frame in the raw backtrace, ensuring that the source line of frame $N$ accurately represents the call site of the callee function in frame $N+1$. For heap-related vulnerabilities, we extract not only the crash backtrace but also the allocation and free backtraces, guiding the LLM to better track the execution states of heap memory objects in subsequent analysis.

Notably, the lack of a sanitizer report does not necessarily preclude \VAA from identifying potential crash backtraces, as long as the target location is provided. Given that many off-the-shelf static analysis tools \cite{codeql, sui2016svf, shi2018pinpoint, kang2022tracer, yao2024falcon} are proficient at source-to-sink path finding, such capabilities can be integrated into our framework. We leave this for future work. % This paragraph is optional.

\smallskip
\parabf{Harness Command Analysis.} Constructing an accurate harness command to trigger the target vulnerability is non-trivial. First, a single codebase may produce multiple CLI binaries, making the mapping between these binaries and the vulnerable code obscure. Second, even when the correct binary is identified, the selection of arguments significantly influences the resulting execution path. We resolves this by identifying potential entry points and locating argument parsing logic and help strings. By analyzing code snippets relevant to the crash backtrace, our agent infers the appropriate harness command for PoV testing. Furthermore, for file-parsing programs, \VAA also extracts necessary input file extensions to ensure the test case can pass the checks. % (e.g., )

\smallskip
\parabf{Root Cause Analysis.} Identifying the root cause (RC) of a vulnerability is fundamental to the PoV generation task, and typically serves as the initial step for manual vulnerability reproduction by security researchers.
While existing literature~\cite{park2024benzene, xu2024racing, yagemann2021arcus, jiang2021igor, yagemann2021automated} leverages symbolic execution or pattern recognition to characterize vulnerability root causes, these methods are often hampered by limited scalability and heavy overhead. Moreover, their outputs typically consist of a set of condition constraints that lack the high-level semantics required for end-to-end PoV synthesis.
In our work, \VAA engages in multi-round interactions with the codebase environment based on the analysis results from the preceding two sub-steps. Once all critical code fragments are incorporated into the context, it delivers a comprehensive analysis of the root cause. 
Specifically, our agent performs root cause analysis from three dimensions: (i) forward reasoning from entry points to derive input format prerequisites; (ii) backward tracing from crash sites to pinpoint the critical conditions that violate security properties; and (iii) deduction from generic vulnerability types to identify instance-specific sub-patterns for the current vulnerability.

After acquiring the comprehensive root cause analysis of a vulnerability, we incorporate these findings as persistent context for all subsequent PoV generation attempts. Equipped with this context, the LLM will no longer reason and solve problems blindly. Instead, it will know which critical contextual dependencies need to be retrieved, thus {mitigating Challenge 1}.  Furthermore, vulnerability analysis insights can also help our agent more intelligently select and examine execution feedback in the latter two phases, thereby improving the task success rate.

\subsection{Code Instrumentation}
\label{subsec:instrumentation}
% 两条自动化流程（也可以讲我们的循环迭代diagnose、fix，以及不同编译系统类型的区分）
%% 输入是basic build script
%% 
% 强调我们有字符串检查的oracles，确保整个系统流程的鲁棒性（提一嘴text wrapper的特别处理）
% 素材：子流程图 ✅（编译与修复的小循环，这个在overview里是没有的）

For a given C/C++ codebase, we perform two distinct types of instrumented compilation. The resulting binaries provide semantic-level execution feedback for the subsequent two phases of \ourtool. 

\begin{figure}[h]
  \centering
  \captionsetup{skip=8pt}
  \includegraphics[width=0.75\linewidth]{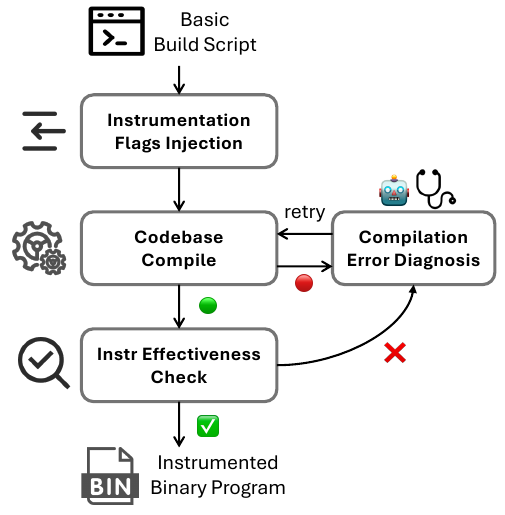} % Value could be adjusted.
  \caption{Automated code instrumentation pipeline.}
  \Description{The workflow of our code instrumentation sub-agent.}
  \label{fig:code_instr}
\end{figure}

In this stage, \ITA first extract the basic build procedures and commands from the project documentation (e.g., README.md). Then, it implements the required instrumentation by injecting corresponding compilation parameters into the basic build script. This design ensures high extensibility across different instrumentation types. With prepared build script, \ITA iteratively attempts to compile the codebase. If compilation errors are encountered, our agent will collect and diagnose the error messages, refine the compilation commands, and retry the process until successful or a predefined attempt limit is reached, as illustrated in Figure \ref{fig:code_instr}.

\begin{itemize}[leftmargin=*]
  % Source-based Code Coverage vs. Edge-based Coverage with Bitmaps
  \item For coverage instrumentation, we inject "{-fprofile-instr-generate}" and "{-fcoverage-mapping}" flags to leverage LLVM's native source-based code coverage profiling. Unlike the bitmap-based coverage feedback commonly employed by fuzzers, this technique enables a more direct and precise mapping between executed code and source lines, which facilitates the reconstruction of semantic-level execution details.
  \item For sanitizer instrumentation, we inject "-fsanitize=<type>" and "-fno-omit-frame-pointer" flags. Leveraging the vulnerability type identified in the previous phase, \ITA selects the most appropriate sanitizer type (e.g., ASan or UBSan). The resulting compiled binary (\FormulaIn{P_{san}}) is critical for the final validation in the end-to-end PoV generation task.
\end{itemize}

% LHY: This paragraph needs improvement (Highlighting Oracle mindset).
It is worth noting that a successful compilation does not always guarantee successful instrumentation. For instance, an LLM agent may insert instrumentation flags into the build scripts; however, these flags may later be overridden by subsequently specified other compilation parameters. 
To resolve this, we introduce lightweight instrumentation effectiveness checks to examine the compiled binary programs. Specifically, we locate the target binary and inspect it for characteristic strings (e.g., "\_\_llvm" or "\_\_sanitizer") that confirm the presence of the intended instrumentation. This design significantly enhances the robustness of the entire analysis process and avoids wasting tokens when no effective execution feedback is available.
% This design significantly enhances the robustness of the entire analysis process, which is consistently adopted across other components of \ourtool. 
For the projects whose entry point is a shell script rather than a binary, we also identify the actual binary path to perform this validation.

% After collecting the basic build procedures and commands from the repository documentation (e.g., README.md), we implement different types of instrumentation by inserting corresponding compilation parameters into the basic build script, which ensures high scalability.

\subsection{Path Exploration}
\label{subsec:path_explore}
% 介绍PEA循环，上一轮生成的testcase会拿给下一轮，在生成新的testcase之前先看下。终止条件的heuristic：最后三个stack frames都被cover到。
% 混合反馈策略：默认情况下，只会把crash trace的粗粒度函数执行情况直接反馈给agent，代码行级的覆盖情况则让agent自己调tool、按需查看。
% PEA uses llvm-cov: 解决path stuck、以到达vulnerable function、满足输入文件格式要求。
% 素材：PEA + CTA 应该有一个一起的算法框图✅。

% Para0: Transition sentences.
Different from existing LLM-based agents that rely on simplistic "generate-and-test" iterations, \ourtool decouples the PoV generation process into two specialized phases: path exploration and crash triggering. The complete workflow of this collaborative process is formalized in Algorithm~\ref{alg:pov-generation}.

% Para1: Introduce general loop from input to end conditions (with algorithm table).
In this phase, \PEA adopts an iterative loop with fine-grained execution feedback, together with a well-designed context maintenance mechanism, to enable execution-state-aware LLM reasoning. 
Specifically, as illustrated in the algorithm~\ref{alg:pov-generation}, \PEA first attempts to generate a test case based on the basic vulnerability information and root cause. Then, the generated test case is executed on \FormulaIn{P_{cov}}, from which \PEA collects program coverage information (using \CodeIn{CollectCov}). After updating the context, \PEA reviews both the test case generated in the previous iteration and the program’s internal execution state before synthesizing the next test case, allowing it to understand the bottlenecks that hinder exploration.

% During this iterative process, we additionally check whether each generated test case can already cover the last three stack frames of the crash backtrace (using \CodeIn{ReachesVulnFunc}). If so, the test case is added to a set of useful test cases (i.e., \CodeIn{useful\_tcs}), which is later leveraged by \CTA to make the PoV meet input file format requirements. % Old version
During this iterative process, we additionally check whether each generated test case already covers the last three stack frames of the crash backtrace (using \CodeIn{ReachesVulnFunc}). If so, the test case is considered useful, as it has reached the vulnerable function with a largely valid execution context, and is added to a set of useful test cases (i.e., \CodeIn{useful\_tcs}). These test cases represent inputs that already pass the program’s parsing and basic format checks. As a result, \CTA can reuse them as a starting point and concentrate on refining inputs to satisfy vulnerability-specific runtime predicates, rather than repeatedly fixing input well-formedness.

% Para2: Introduce how we bridge the semantic gap in execution feedback (address Challenge2). The hybrid feedback strategy. 
% The tools we use and the goal of this phase.
Although the \texttt{coverage\_query} agent tool built upon llvm-cov is capable of retrieving coverage information at various granularities (e.g., line-level and region-level), guiding an LLM to utilize this capability both stably and effectively remains a significant challenge. 
To address this, we propose a \emph{hybrid feedback strategy} to balance pre-defined contextual information with the autonomy of the LLM agent. On the one hand, during \CodeIn{CollectCov}, \PEA deterministically collects and injects coarse-grained, function-level coverage information along the crash backtrace into the context. This step does not require the LLM to generate any tool calls. On the other hand, \PEA can autonomously query fine-grained, line-level coverage information on demand by explicitly invoking the \CodeIn{coverage\_query} tool and inspecting its results. 
This design allows the LLM to identify where path exploration becomes stalled using a appropriate number of tool calls, enabling more targeted test-case refinement.
Ultimately, this hybrid feedback strategy bridges the semantic gap between low-level execution states and high-level LLM reasoning, thereby addressing Challenge 2.

% Collaboration between PEA and CTA
% Path Exploration and Crash Triggering
\begin{algorithm}[h]
  \caption{Collaboration between \PEA and \CTA}
  \label{alg:pov-generation}
  \begin{algorithmic}[1]
  \REQUIRE Vulnerability info $\mathcal{V}$, Crash backtrace $\mathcal{T}$, Root cause $\mathcal{R}$
  , \\ Instrumented binary programs $P_{cov}$ and $P_{san}$
  \ENSURE Proof-of-Vulnerability $\mathcal{P}$
  \vspace{0.4em}

  \STATE \textbf{/* Path Exploration Phase */}
  \STATE $useful\_tcs \gets \emptyset$ %, $context \gets \emptyset$
  \STATE $context \gets \textsc{UpdateContext}(\mathcal{V}, \mathcal{R})$
  \FOR{$i = 1$ \TO $N_1$}
      \STATE $tc \gets \textsc{LLM-Generate-Testcase}(context)$
      \STATE $raw\_profile \gets \textsc{Execute}(P_{cov}, tc)$ 
      \STATE $cov\_info \gets \textsc{CollectCov}(raw\_profile)$
      \IF{$\textsc{ReachesVulnFunc}(cov\_info, \mathcal{T})$}
          \STATE $useful\_tcs \gets useful\_tcs \cup \{tc\}$
      \ENDIF
      \STATE $context \gets \textsc{UpdateContext}(tc, cov\_info)$
  \ENDFOR
  \vspace{0.5em}

  \STATE \textbf{/* Crash Triggering Phase */}
  \STATE $vuln\_type, guidance \gets \textsc{SampleVulnTypeHints}(\mathcal{R})$
  \STATE $context \gets \textsc{UpdateContext}(\mathcal{V}, \mathcal{R}, guidance, useful\_tcs)$
  \FOR{$j = 1$ \TO $N_2$}
      \STATE $pov \gets \textsc{LLM-Generate-PoV}(context)$
      \STATE $exec\_info \gets \textsc{Execute}(P_{cov}, pov)$
      \IF{$\textsc{ValidateCrash}(P_{san}, pov) = vuln\_type$}
          \RETURN $pov$
      \ENDIF
      \STATE $context \gets \textsc{UpdateContext}(pov, exec\_info )$
  \ENDFOR

  \RETURN \texttt{null}
  \end{algorithmic}
\end{algorithm}

\subsection{Crash Triggering}
\label{subsec:crash_trigger}
% 介绍CTA的循环与引导，还有prompt sampler的设计。
% CTA uses llvm-cov: 解决triggering conditions的stuck，写出一些非常规代码。
% 素材：prompt sampler的引导提示词片段。

% Para1: Introduce general loop from input to end conditions (with algorithm table). Especially the crash triage/validation operation.
In the crash triggering phase, \CTA takes as input the accumulated useful test cases from path exploration and iteratively attempts to construct a PoV that can trigger the target crash. 
As formalized in Algorithm~\ref{alg:pov-generation}, this phase follows a generate–execute–validate loop. In each iteration, \CTA synthesizes a PoV candidate from the current context, execute it to observe runtime behavior, and then validated against the sanitizer-instrumented binary (using \CodeIn{ValidateCrash}).
If a PoV candidate produces a sanitizer error consistent with the ground-truth crash, the process terminates successfully. Otherwise, execution feedback is incorporated into the context to guide subsequent refinement, allowing \CTA to progressively move from near-crashing executions toward a valid PoV.

% Para2: Introduce the prompt sampler (address Challenge 3), how we guide LLM to trigger the vuln crash according to different vuln types.
A key challenge is that the content of a PoV often differs substantially from that of regular input files, making it difficult for the LLM to deviate from its pretraining data and safety-oriented post-training.
To mitigate this issue ({Challenge 3}), \CTA leverages a \emph{lightweight prompt sampler} that derives vulnerability-type-specific guidance from the root cause, explicitly steering the LLM toward inputs that are more likely to satisfy the vulnerability-triggering runtime predicates. 
Concretely, \CodeIn{SampleVulnTypeHints} triages the root cause to identify the vulnerability type and then provides corresponding high-level hints that bias the LLM’s reasoning toward inputs that are more likely to violate the associated security properties.
% Rather than hard-coding triggering patterns, these hints serve as soft constraints that steer the LLM’s generation process while preserving its flexibility to explore unconventional triggering paths. 
This enables our agent to adapt its generation strategy across different vulnerability classes without relying on hand-crafted exploit templates.
% \revision{maybe add a prompt sampler example...} 

% Para3: The tools we use and the different goal of llvm-cov in this phase (inspect and solve constraints to triggering the vuln).
It is worth noting that the role of execution feedback during crash triggering is different from the path exploration phase. 
Although the \texttt{coverage\_query} tool remains available, coverage information is primarily used to help analyze why a PoV candidate fails to trigger the crash, rather than to expand exploration coverage. 
By selectively querying fine-grained coverage data and correlating it with sanitizer feedback, \CTA can focus its reasoning on the final obstacles to crash triggering, resulting in more directed and efficient PoV refinement.

%% file: sections/4_implementation.tex
\section{Implementation}
\label{sec:implementation}

We built \ourtool from scratch, resulting in a codebase of approximately 7,700 lines of Python code. 
While several off-the-shelf CLI-based coding agents (such as Cursor CLI~\cite{cursorcli} and OpenAI Codex~\cite{codex}) allow for customization, recent research~\cite{xia2024agentless, xiao2024flowbench, shi2025flowagent, yang2025harnessagent} suggests that agents relying solely on LLM-driven autonomous planning often suffer from prompt drift and performance instability when tackling complex tasks.
% \textbf{Engineering Philosophy.} 
In our implementation, we adopt a pre-structured planning approach, where the four primary phases are executed sequentially in a fixed order. The overall workflow is designed to emulate the common operating procedures followed by human security experts when constructing a PoV. Within each phase, we deploy a dedicated sub-agent that maintains a high degree of autonomy over decision-making and tool invocation within a stabilized work cycle. 
This hybrid design strikes a strategic balance between the reliability of structured workflows and the generalization capabilities inherent in autonomous agents.

\textbf{Agent Tools.} To facilitate interaction with the target codebase and execution environment, \ourtool provides a comprehensive suite of tools, namely \CodeIn{read\_file}, \CodeIn{write\_file}, \CodeIn{execute\_bash}, \CodeIn{finish}, \CodeIn{ast\_grep\_search\_function}, \CodeIn{ast\_grep\_search\_pattern}, and \CodeIn{coverage\_query}. 
%% Emphasize the cov tool %%
Specifically, the coverage query tool is built upon the "llvm-cov" and "llvm-profdata" commands. It parses the "default.profraw" files generated after each execution to extract fine-grained, source-mapped coverage information and presents it in an agent-friendly format.
%% Python capability explain %%
Furthermore, our agent’s capability to effectively construct structured binary input data (such as MP4 video streams) is derived from a two-step process: first, it utilizes \CodeIn{write\_file} to generate a Python script containing the logic for synthesizing structured binary bytes; subsequently, it invokes \CodeIn{execute\_bash} to run the script, thereby producing the final test case or PoV candidate file.
% Notably, Python execution is achieved through the composition of \texttt{write_file} and \texttt{execute_bash}. The agent can first write python scripts to disk, then execute them via bash command \texttt{python3}, which maintains interface simplicity while supporting complex file/data manipulation. 

% Runtime Cost Control/Mitigation
\textbf{Runtime Cost Control.}
Many prior studies~\cite{titanfuzz, zeng2025pbfuzz, jain2025testforge} report that directly using LLM-based agents for test case generation incurs substantial monetary overhead, particularly when generating complex PoVs. In the implementation of \ourtool, we adopt several techniques to mitigate the runtime cost.
First, \ourtool supports flexible configuration of different LLMs for different sub-tasks. For example, root-cause analysis is performed using a more capable model (e.g., Claude-Sonnet), while crash trace refinement is delegated to a more lightweight and cost-efficient model (e.g., Claude-Haiku). This design optimizes the trade-off between analysis quality and API cost.
Second, we impose explicit length limits on outputs returned by each tool invocation, preventing context explosion and reducing unnecessary token consumption.
Finally, \ourtool continuously monitors the cumulative cost during analysis and terminates the process once a predefined budget threshold is exceeded.
Together, these design choices make our purely LLM-agent-based approach practical and cost-effective in real-world settings.

% \textbf{Oracle Mindset.}
%% 1. instrumentation check
%% 2. sanitizer-based crash check

%% file: sections/5_evaluation.tex
\section{Evaluation}
% OpenHands + sonnet 4.5:
%% 1. no budget limit: 60 (id. 01~60)
%% 2. budget limit \$1.5: 15 (id. 01~75)
% OpenHands + Gpt-5.2:
%% 1. budget limit \$1.5: Abandoned due to unknown bug.

In this section, we evaluate \ourtool through the following research questions:

\begin{itemize}[leftmargin=*] %, itemsep=4pt
  % Source-based Code Coverage vs. Edge-based Coverage with Bitmaps
  \item \textbf{RQ1:} How effective is \ourtool in generating validated PoVs for real-world C/C++ vulnerabilities? 
  \item \textbf{RQ2:} What is \ourtool's runtime overhead? % Compare with unlimited budget openhands+sonnet-4.5
  \item \textbf{RQ3:}  Is root cause analysis critical for our agent to retrieve contextual dependencies efficiently?
  \item \textbf{RQ4:}  How does fine-grained execution coverage feedback affect \ourtool's performance? % Illustrate with case study and drillagent.log analysis.
  \item \textbf{RQ5:} Is it necessary to decompose PoV generation into two phases (i.e., path exploration and crash triggering)?
\end{itemize}

\subsection{Experimental Setup}
% \revision{LHY: We don't need to say how many rounds we set ($N_1 and N_2$ in algorithm), it is heuristically adjusted.}

% 可能还得提一嘴SEC-bench里有些tasks的形式是有点问题的（畸形）（gpac.cve-2023-42298的sanitizer report）
\parabf{Benchmarks.}
We evaluate our approach on SEC-bench~\cite{lee2025sec}, a recent large-scale benchmark of real-world C/C++ CVEs designed for software security tasks.
Compared with Magma~\cite{hazimeh2020magma}, SEC-bench covers a more diverse set of open-source projects (29 versus 7), which enables a better assessment of the generalization capability of our approach. 
Besides, SEC-bench provides ready-to-use basic build scripts and raw harness commands, allowing the agent to focus more on the PoV generation task itself.

The original SEC-bench PoV-generation suite contains 200 CVE tasks.
% We exclude 10 tasks whose sanitizer reports provide only raw binary addresses without source-line information in the crash backtraces.
We remove 10 tasks whose sanitizer reports contain only raw binary addresses in crash backtraces and lack source-line information, making them unsuitable for reliable evaluation.
Based on the remaining tasks, we construct two benchmark sets used throughout this paper:

\begin{itemize}[leftmargin=*]
  \item \textbf{SEC-bench-Full (190 tasks).}
  The full evaluation set after filtering the mentioned 10 tasks.
  We report \ourtool's end-to-end PoV generation performance on this set to characterize overall effectiveness in a comprehensive manner.
  \item \textbf{SEC-bench-60 (60 tasks).}
  A uniformly random subset sampled from SEC-bench-Full using a fixed random seed (42), whose corresponding task IDs will be released as part of our artifact.
  We use this subset for head-to-head comparisons with baselines to control evaluation cost while preserving a fair, fixed task set for reproducibility.
  % The seed is in evaluation/utils/shared.py:254-263. The prepare_dataset function uses a hardcoded random seed of 42 via dataset.sample(..., random_state=42, replace=False).
\end{itemize}

% OpenHands. SWE-agent (maybe)
\smallskip
\parabf{Baselines.}
We compare \ourtool against \textbf{OpenHands}~\cite{wang2024openhands}, a state-of-the-art LLM-based agent framework that demonstrates strong performance on SEC-bench.
OpenHands represents a general-purpose agentic system capable of repository navigation, code inspection, tool invocation, and iterative reasoning, and has been shown to be effective on a wide range of software engineering and security tasks.
We select OpenHands as our primary baseline as it embodies the dominant end-to-end paradigm for PoV generation, serving as a representative contrast to our task-specialized, feedback-driven design.

\smallskip
\parabf{Model Configuration.}
To ensure a fair comparison, we evaluate both \ourtool and the baseline agent using the same underlying LLM whenever applicable.
Specifically, unless otherwise noted, we run both systems with \textbf{Claude Sonnet 4.5}, replacing the originally reported model used in OpenHands.
This setup isolates the effect of agent design from differences in model capability.
Within \ourtool, we additionally allow different sub-agents to employ different LLM variants for cost–quality trade-offs (Section~\ref{sec:implementation}), but this configuration is fixed across all evaluation runs and does not affect baseline comparisons.

\smallskip
\parabf{Metrics.}
Our evaluation follows the PoV validation protocol defined by SEC-bench, which determines PoV validity based on sanitizer-reported crash outputs under the benchmark-provided execution configuration.
Instead of treating PoV generation as a binary success-or-failure task, we explicitly distinguish between different classes of all sanitizer-triggering test cases. Given an input that produces a sanitizer-reported crash, we categorize it as one of the following:
\begin{itemize}[leftmargin=*]
  \item \textbf{Validated PoV.}
  An input whose execution triggers a sanitizer crash that matches the \emph{ground-truth vulnerability specification} provided by SEC-bench, including both the expected crash type and the target crash location.
  % This corresponds to a faithful reproduction of the reported vulnerability.
  \item \textbf{Variant PoV.}
  An input that triggers a sanitizer-reported crash but does \emph{not} match the ground-truth vulnerability type or location.
  Such inputs may expose other related memory safety violations or alternative crash sites, which also carry substantial implications for real-world security practices.
\end{itemize}
We record both validated and variant PoVs generated by \ourtool.
Validated PoVs represent faithful reproductions of the target vulnerabilities, while variant PoVs serve as a complementary signal of the agent’s ability to explore and trigger diverse sanitizer-reported failure modes.
% Accordingly, our evaluation reports the absolute number of tasks for which validated and variant PoVs are generated, rather than aggregating results into a single success-rate metric.

\subsection{RQ1: Effectiveness}
\label{subsec:rq1}

\begin{table*}[t]
  \centering
  \caption{End-to-End PoV Generation Effectiveness (RQ1)}
  \label{tab:effectiveness}
  \small
  \begin{threeparttable}
    % \begin{tabular}{l l l c c c c c}
    % \begin{tabular*}{\textwidth}{@{\extracolsep{\fill}} l l l c c c c c}
    \begin{tabular*}{0.85\textwidth}{@{\extracolsep{\fill}} l l l c c c c c}
      \toprule
      \textbf{Method} & \textbf{Benchmark} & \textbf{Budget} &
      \textbf{Validated} & \textbf{Variant} & \textbf{Total} & \textbf{Resolved} & \textbf{Crash}\\
      & & &
      \textbf{PoVs} & \textbf{PoVs} & \textbf{Crashes} & \textbf{Rate$^\dagger$} & \textbf{Rate$^\dagger$}\\
      \midrule
      DrillAgent & SEC-bench-Full & \$1.5/task & \textbf{55} & 12 & \textbf{67} & 28.9\% & \textbf{35.3\%} \\
      \midrule
      DrillAgent & SEC-bench-60   & \$1.5/task & 15 & 2 & 17 & 25.0\% & 28.3\% \\
      OpenHands  & SEC-bench-60   & \$1.5/task &  6 & \textendash &  6 & 10.0\% & 10.0\% \\
      % \midrule
      OpenHands  & SEC-bench-60   & \small{unlimited}    & 18 & \textendash & 18 & {30.0\%} & 30.0\% \\
      \bottomrule
    \end{tabular*}

    % SOTA overall resolved rate , note. Obvious.
    \begin{tablenotes}[flushleft]
      \footnotesize
      \item[$\dagger$] Resolved Rate is computed as the number of Validated PoVs divided by the total number of tasks in the benchmark. 
      Crash Rate is computed as the total number of crashes (i.e., Validated PoVs + Variant PoVs) divided by the total number of tasks.
      % Resolved Rate is calculated as \textit{Validated PoVs} divided by the total number of tasks in the current benchmark.
      % \item Our number of \textit{Validated PoVs} is 52.8\% higher than the current state-of-the-art result~\cite{secbsite}.
    \end{tablenotes}
  \end{threeparttable}
\end{table*}

Table~\ref{tab:effectiveness} summarizes the end-to-end PoV generation effectiveness of \ourtool and the baseline agent OpenHands on SEC-bench.
Overall, \ourtool demonstrates a clear advantage in generating validated Proofs-of-Vulnerability under a fixed budget constraint.

\parabf{Overall PoV Generation Performance.}
On the full benchmark (SEC-bench-Full), \ourtool successfully generates validated PoVs for 55 out of 190 tasks, achieving a resolved rate of 28.9\%. 
The current highest reported result on the SEC-bench leaderboard~\cite{secbsite} achieves an 18\% resolved rate (36 solved tasks). Compared to this result, \ourtool improves the resolved rate from 18\% to 28.9\%, corresponding to \textbf{a 52.8\% relative increase in the number of solved CVE tasks (55 vs. 36)}.
% As of now, the highest reported resolved rate on the SEC-bench leaderboard~\cite{secbsite} is 18\% by OpenHands + Claude Sonnet 3.7, corresponding to 36 solved tasks. In comparison, \textbf{\ourtool improves upon this state-of-the-art result by 52.8\%}. 
Moreover, among the 55 tasks that \ourtool successfully solves and validates, 29 are uniquely solved (i.e., they have not been successfully addressed by any prior methods or evaluations). This result highlights the distinct advantages and complementary strength of our approach.

In addition to the 55 validated PoVs, \ourtool also generated 12 {variant PoVs}, which mainly trigger sanitizer-detected crashes with different error types (e.g., producing a heap buffer overflow instead of the originally expected memory leak).
Although these variant PoVs are not counted as successful reproductions under the SEC-bench evaluation protocol, they demonstrate the agent’s ability to reach semantically relevant program states and uncover related failure patterns. 
To capture this capability beyond strict vulnerability resolution, we additionally report the \textit{Crash Rate}, on which \ourtool achieves an impressive 35.3\%.
Such PoV variants also carry practical security value: in the AIxCC competition~\cite{AIxCC2023}, variants that trigger the target vulnerability via different crash paths can receive additional credit, as they may bypass specific patches.

\parabf{Comparison with OpenHands under Controlled Budget.}
On the SEC-bench-60 subset, \ourtool consistently outperforms the OpenHands baseline under the same Claude Sonnet 4.5 model and budget constraint of \$1.5 per task.
Specifically, our approach generates more than twice as many validated PoVs as OpenHands (15 vs. 6), highlighting the effectiveness of its execution state feedback design.
Notably, when the budget constraint is removed, OpenHands achieves a higher absolute success rate; however, this comes at the cost of unbounded runtime and monetary overhead, which will be discussed in the next section.
% In contrast, \ourtool is explicitly designed to operate within a strict budget while still maintaining competitive effectiveness.

\parabf{Breakdown of Solved Tasks and Execution Behavior.} 
To gain a deeper understanding of the success of \ourtool, we conduct a more fine-grained statistical analysis over the evaluation results.

Among the 67 tasks on which \ourtool successfully triggered crashes, these tasks span 13 unique C/C++ projects, nearly half of the 29 projects included in SEC-bench, demonstrating that the success is not concentrated on a few specific codebases. 
The generated PoVs cover a wide spectrum of input formats: some require syntactically valid programming language scripts (e.g., mruby and njs), while others involve highly structured binary multimedia files (e.g., gpac and imagemagick). This diversity demonstrates the strong generalization capability of our approach across heterogeneous application domains and input formats.  

\begin{figure}[h]
  \centering
  \captionsetup{skip=8pt}
  \includegraphics[width=0.94\linewidth]{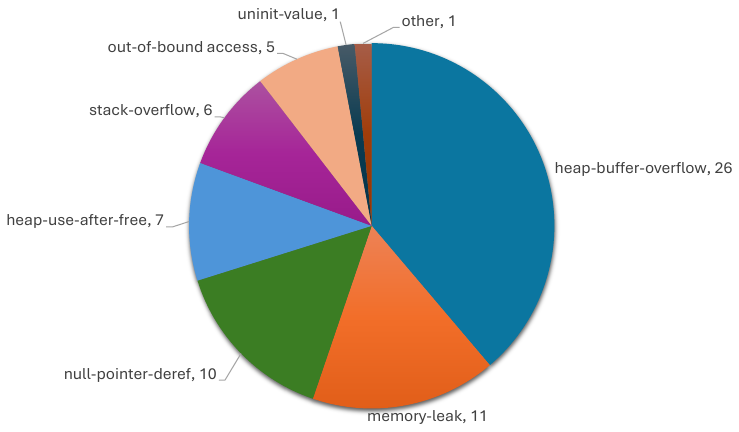}
  \caption{Vulnerability-Type Distribution of Solved Tasks.}
  \Description{Vulnerability-Type Distribution of Solved Tasks.}
  \label{fig:vuln_types}
\end{figure}

As shown in Figure~\ref{fig:vuln_types}, the types of vulnerabilities successfully handled by \ourtool also broadly cover those present in the benchmark, including stack overflow, heap buffer overflow, use-after-free, null pointer dereference, and memory leak. This indicates that the effectiveness of \ourtool is not limited to a specific vulnerability category, but extends to a wide range of memory safety issues.  
% TODO: 最好再说一下variant PoVs的漏洞类型映射情况。

\begin{figure}[h]
  \centering
  \captionsetup{skip=8pt}
  \includegraphics[width=0.80\linewidth]{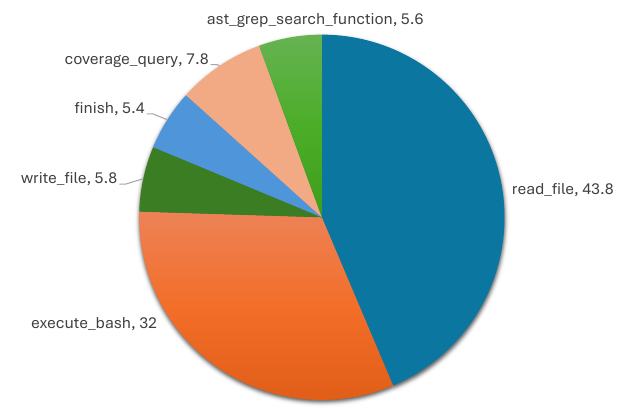}
  \caption{Average Tool Call Count per Task.}
  \Description{Average Tool Call Count per Task.}
  \label{fig:tool_usage}
\end{figure}

Furthermore, to better understand the execution behavior of \ourtool, we analyze its tool usage over all 190 tasks. On average, each end-to-end task requires 60 LLM interactions and 100.4 tool invocations. The detailed distribution of tool call types is illustrated in Figure~\ref{fig:tool_usage}.  
We observe that the majority of tool calls are devoted to source code reading, which is consistent with the fact that the input tokens consumed in the PoV generation task far exceed the output tokens (Table~\ref{tab:phase_breakdown}). 
% This observation aligns with Challenge~1, where contextual dependency retrieval dominates the reasoning cost in PoV generation.
Although the \CodeIn{coverage\_query} tool is invoked only 7.8 times per task on average, this number is biased downward by tasks that generate a PoV in a single attempt or fail during instrumented compilation. In practice, coverage feedback is crucial for the success of many tasks, as will be further demonstrated in Section~\ref{sec:rq4}.
% Furthermore, to better understand the execution behavior of \ourtool, we analyze its tool usage over all 190 tasks. On average, each end-to-end task requires 60 LLM interactions and 100.5 tool invocations. The detailed distribution of tool call types is illustrated in Figure~\ref{fig:tool_usage}.

% Statistics
% Tool call num and tool type distribution. %实现的稳定性
% Solved tasks distribution (what projects, what vuln types). %方法的泛化性
% 这部分数据就只在effectiveness加一段话提一嘴就行，因为看起来其实没有超过history的泛化性（当然openhands+swe-agent两个比我一个也不公平orz）
% TODO: Current tool call success rate is 616/1481 (41.6%). Since it is kind of low, we just skip mentioning it.

\parabf{Summary Findings.}
Overall, \ourtool substantially improves end-to-end PoV generation effectiveness under fixed budget constraints. The results on SEC-bench demonstrate that our approach not only outperforms a strong LLM-agent baseline, but also advances the current best resolved rate while reproducing a considerable number of previously unresolved vulnerabilities. 
% The task-level analysis further shows that the success of \ourtool generalizes across diverse projects, input formats, and vulnerability types, rather than being concentrated on a narrow subset of cases. Moreover, the behavioral statistics reveal that execution feedback plays a central role in guiding the agent’s reasoning process, enabling it to iteratively bridge the gap between semantic hypotheses and concrete runtime constraints.
% These findings collectively highlight the importance of execution-state-aware feedback for reliable and scalable PoV reproduction.
These gains highlight the importance of execution-state-aware feedback for reliable and scalable PoV reproduction.
% , rather than model capacity alone, 

\subsection{RQ2: Efficiency}
\label{subsec:rq2}

We first analyze the monetary cost of our approach, as shown in Table~\ref{tab:efficiency}.
On SEC-bench-Full, \ourtool incurs an average cost of \$1.79 per task under a nominal budget of \$1.5/task. This slight overrun is expected, as \ourtool does not enforce a hard cutoff at the budget limit; instead, it guarantees the completion of at least one full test-case generation and validation cycle to avoid premature termination. Despite this conservative design choice, the overall cost remains tightly controlled. 
On the SEC-bench-60 subset, \ourtool exhibits comparable behavior, with an average cost of \$1.93 per task, remaining close to the intended budget while substantially outperforming the OpenHands baseline in terms of validated PoV generation.

A more revealing comparison emerges by calculating the average cost spent on each successfully resolved task.
On SEC-bench-60, {\ourtool achieves a cost of \$7.72 per validated PoV}, which is significantly lower than OpenHands under both budgeted (\$15.30) and unlimited (\$20.13) settings. Although OpenHands solves slightly more tasks when allowed unlimited budget, this improvement comes at a disproportionate monetary cost. In contrast, \ourtool delivers a substantially better cost–effectiveness trade-off, demonstrating that execution-state-aware reasoning enables more efficient convergence toward valid PoVs rather than relying on unconstrained exploration.

We next examine execution time overhead.
As shown in Table~\ref{tab:efficiency}, \ourtool requires approximately 11–12 minutes per task, which is higher than OpenHands but remains within a practical range for automated vulnerability reproduction. Table~\ref{tab:phase_breakdown} further indicates that this time is distributed across multiple reasoning-intensive phases. Importantly, this overhead is modest compared to traditional fuzzing-based approaches, which often require hours of execution to reproduce deep, logic-heavy vulnerabilities. For such cases, our method offers a more time-efficient alternative while maintaining deterministic PoV validation.

% Overall, DrillAgent achieves a favorable balance between cost, time, and effectiveness.
% While incurring moderate additional overhead per task, it significantly reduces the monetary cost per successful PoV and avoids the unbounded expenses associated with unrestricted agentic exploration. These results confirm that DrillAgent is not only effective but also efficient for budget-aware, real-world PoV generation.

% Need to include: Average time and cost (input/output tokens). Max/Min value. (different phases and total) (In which phase or which round, the PoV is generated?) $\checkmark$
\begin{table*}[t]
  \centering
  \caption{End-to-End Task Cost and Execution Time (RQ2)}
  \label{tab:efficiency}
  \small
  \begin{threeparttable}
    % \begin{tabular}{l l l c c c}
    \begin{tabular*}{0.85\textwidth}{@{\extracolsep{\fill}} l l l c c c}
      \toprule
      \textbf{Method} & \textbf{Benchmark} & \textbf{Budget} &
      \textbf{Cost/Task} & \textbf{Cost/Success*} & \textbf{Exec Time} \\
      & & &
      \textbf{(\$)} & \textbf{(\$)} & \textbf{(min)} \\
      \midrule
      DrillAgent & SEC-bench-Full & \$1.5/task & 1.79 & \textbf{6.18} & 11.5 \\
      \midrule
      DrillAgent & SEC-bench-60   & \$1.5/task & 1.93 & \textbf{7.72} & 11.8 \\
      OpenHands  & SEC-bench-60   & \$1.5/task & 1.53 & 15.30 & 3.2 \\
      OpenHands  & SEC-bench-60   & \small{unlimited}  & 6.04 & 20.13 & 7.7 \\
      \bottomrule
    \end{tabular*}

    \begin{tablenotes}[flushleft]
      \footnotesize
      \item *Cost / Success is computed as the total monetary cost divided by the number of validated PoVs. 
      {And the three above metrics are averaged over all tasks.}
    \end{tablenotes}
  \end{threeparttable}
\end{table*}

\begin{table*}[t]
  \centering
  \caption{Phase-wise Time and Token Usage Breakdown of \ourtool on SEC-bench-Full}
  \label{tab:phase_breakdown}
  \small
  \begin{threeparttable}
    % \begin{tabular}{l c c c}
    \begin{tabular*}{0.65\textwidth}{@{\extracolsep{\fill}} l c c c}
      \toprule
      \textbf{Phase} &
      \textbf{Avg. Time (min)} & %\textbf{Cost} &
      \textbf{Input Tokens} & \textbf{Output Tokens} \\
      \midrule
      Vulnerability Analysis & 2.5 &  %\textit{0.33} & 
      \textit{228,145} & \textit{8,147} \\
      Code Instrumentation        & 1.9 &  %\textit{-} & 
      \textit{-*} & \textit{-*} \\
      Path Exploration       & 3.7 &  %\textit{0.78} & 
      \textit{195,908} & \textit{12,748} \\
      Crash Triggering        & 3.3 &  %\textit{0.68} & 
      \textit{171,438} & \textit{11,584} \\
      % \midrule
      % \textbf{Total}              & 11.5 & 1.79 & \textit{367,346} & \textit{24,333} \\
      \bottomrule
    \end{tabular*}

    \begin{tablenotes}[flushleft]
      \footnotesize
      % \item Results are aggregated over all tasks in SEC-bench-Full.
      \item *Token usage during the code instrumentation phase is omitted, as it involves few LLM interactions and no tool calls.
    \end{tablenotes}
  \end{threeparttable}
\end{table*}

\subsection{RQ3–RQ5: Ablation Study}
% \subsection{Ablation Study}
\label{subsec:ablation}
% LHY: After ablating certain design, we only compare the overall resolved rate.
To understand which components are critical to \ourtool's effectiveness, we conduct an ablation study that systematically disables individual mechanisms in the full system.
Considering the cost constraints, we perform ablation experiments on \textbf{SEC-bench-Ablation-30}, a carefully selected 30-task subset of SEC-bench-Full containing only tasks successfully solved by \ourtool.
To maximize the diagnostic value of the ablation results, we prioritize tasks that are uniquely solved by \ourtool, as these cases better highlight the contributions of individual design components.
% 可以提一嘴这30个tasks里面有多少是被 DrillAgent uniquely sovled，21个。
% 仅在 ablation_subset.txt 中的 9 个 task（不在 unique_validated_pov_set 中）：
%   - exiv2 的 3 个（cve-2017-18005, cve-2018-17230, cve-2020-18899）
%   - imagemagick 的 2 个（cve-2018-5248, cve-2022-48541）
%   - libplist.cve-2017-5545
%   - openjpeg 的 2 个（cve-2016-10507, cve-2017-14041）
%   - upx.cve-2017-15056 

% We compare the full \ourtool system with the following ablated variants:
% \textbf{w/o RCA}, \textbf{w/o ExecFB}, and \textbf{w/o CrashGuide}.
% Importantly, all ablated variants are evaluated on the \emph{same fixed task set}, ensuring that observed performance differences are attributable to system design changes rather than task variance.
% Considering the cost constraints, we select 30 CVE tasks from the validated PoVs successfully generated by \ourtool, focusing on those with unique solutions or longer crash traces, and conduct ablation experiments on them to investigate the factors contributing to \ourtool’s superior performance. 

\subsubsection{Impact of Root Cause Analysis (RQ3)}
\label{sec:rq3}
To evaluate the role of root cause analysis (RCA), we remove all root-cause-related context from \ourtool and only retain the crash location and basic vulnerability information. As shown in Figure~\ref{fig:ablation}, disabling RCA leads to a clear reduction in the number of validated PoVs, indicating that RCA is critical for effective PoV generation.

This result suggests that, for PoV synthesis, high-level root cause descriptions are more effective than rigidly collecting all low-level program constraints. RCA enables the agent to focus on semantically relevant contextual dependencies and guides efficient context retrieval, thus mitigating Challenge 1. This also partially explains why prior approaches such as FaultLine~\cite{nitin2025faultline} exhibit weaker performance on C/C++ programs, where constraint-heavy reasoning without high-level abstraction becomes brittle and impractical.

\subsubsection{Impact of Execution State Feedback (RQ4)}
\label{sec:rq4}
To study the impact of execution state feedback, we disable all coverage query capabilities in \PEA and \CTA, making the agent execution-state-unaware. As illustrated in Figure~\ref{fig:ablation} (refer to \CodeIn{w/o ExecFB}), this ablation also causes severe performance drop.

Without execution feedback, the agent actually reasons “blindly” and lacks actionable signals to understand why a generated input fails. As a result, it becomes difficult to simultaneously satisfy path reachability and vulnerability-triggering constraints, confirming that execution-state-aware feedback is essential for grounding LLM reasoning in concrete program behavior.

% Claude code prompt for ablation code
% 我要在目前这一版本drill agent的基础上做消融实验，具体消融的对象是src/drillagent/tools/cov_query_tools.py和 /home/ubuntu/PoV/DrillAgent/src/drillagent/utils/llvm_cov.py里面的内容。
% 具体来说，对于PathExplorationAgent，_collect_coverage_feedback的操作直接不需要了，_analyze_coverage_feedback里只有BASIC EXECUTION INFO需要被反馈给LLM，_generate_test_case里面的run_agent_loop也不允许提供FunctionCovTool这一工具了。
% 对于CrashTriggerAgent，主要是_generate_pov_candidate里面的run_agent_loop也不允许提供FunctionCovTool这一工具了。
% 此外，src/drillagent/prompts/PathExplorationAgent_prompt.py 和src/drillagent/prompts/CrashTriggerAgent_prompt.py里面的提示词也需要做对应修改，即不再引导LLM查看覆盖率信息或进行coverage_query的工具调用。
% 请你帮我修改上述内容，有不懂的点可以先问我。

\subsubsection{Impact of Crash-Triggering Guidance (RQ5)}
\label{sec:rq5}
\begin{figure}[h]
  \centering
  \captionsetup{skip=6pt}
  \begin{tikzpicture}
    \begin{axis}[
        width=0.85\columnwidth,
        height=0.50\columnwidth,
        ybar,
        bar width=12pt,
        ymin=0,
        ymax=38,
        enlarge x limits=0.12,
        ylabel={Number of PoVs}, %%%%%%%%%
        symbolic x coords={
            DrillAgent-Full,
            w/o RCA,
            w/o ExecFB,
            w/o CrashGuide
        },
        xtick=data,
        xticklabel style={
            rotate=20,
            anchor=east,
            font=\small
        },
        ymajorgrids=true,
        grid style=dashed,
        legend style={
            at={(0.5,1.03)},
            anchor=south,
            legend columns=2,
            font=\small
        },
        nodes near coords,
        nodes near coords style={
            font=\scriptsize,
            yshift=2pt
        }
    ]

    % -------- Validated PoVs --------
    \addplot coordinates {
        (DrillAgent-Full,    30)
        (w/o RCA,            15)
        (w/o ExecFB,         17)
        (w/o CrashGuide,     20)
    };

    % -------- Variant PoVs --------
    \addplot coordinates {
        (DrillAgent-Full,    0)
        (w/o RCA,            3)
        (w/o ExecFB,         2)
        (w/o CrashGuide,     2)
    };

    \legend{Validated PoVs, Variant PoVs}

    \end{axis}
  \end{tikzpicture}
  \caption{Ablation study on key \ourtool design components. Each variant disables one component while keeping all other settings fixed.}
  \label{fig:ablation}
\end{figure}
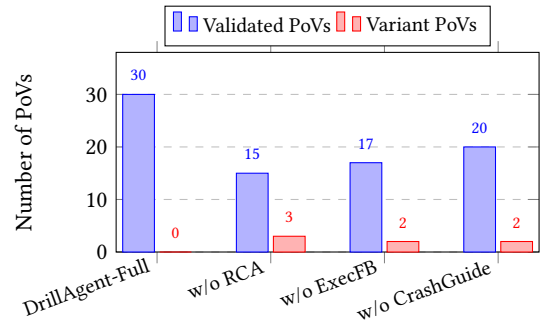

For this research question, we remove the dedicated crash-triggering guidance by disabling \CTA and integrating part of its functionality into \PEA. As shown in Figure~\ref{fig:ablation}, the resolved rate decreases, but the degradation is less severe compared to other ablations.

This indicates that while explicit crash-triggering guidance is not strictly required for reachability, it plays an important role in the final vulnerability-triggering stage. In particular, the prompt sampler mentioned in Section~\ref{subsec:crash_trigger} helps bias the LLM toward security-violating behaviors, alleviating the difficulty of triggering crashes beyond mere path exploration.

% Claude code prompt for ablation code
% 我要在目前这一版本drill agent的基础上做消融实验，具体消融的对象是: src/drillagent/pov_analysis/crash_triggering.py
% 具体来说，我想要实验测试：假如没有CrashTriggerAgent和class VulnTypeGuidanceSampler这么翔实的、关于漏洞触发的引导，DrillAgent的表现还有这么好吗？
% 请你移除整个CrashTriggerAgent的运行，但与之相应地，你需要把CrashTriggerAgent没有跑的self.max_trigger_iterations和self.max_generation_turns 的轮次，加到 src/drillagent/pov_analysis/path_exploration.py的PathExplorationAgent的init函数相应成员，以确保迭代生成的总轮数保持一样多。 也请你对src/drillagent/prompts/PathExplorationAgent_prompt.py里面的系统提示词和任务提示词进行少量的修改，将LLM的目的从“到达漏洞函数”改为也要尝试违反内存安全规则、触发漏洞崩溃，可以参考src/drillagent/prompts/CrashTriggerAgent_prompt.py里面的一些表述，但不用说的那么详细，带一点这个意思就行了。
% 请你帮我修改上述内容，有不懂的点可以先问我。

%src/drillagent/pov_analysis/crash_triggering.py里面有其独有的_record_final_pov和_record_variant_pov，其会在self.drill_agent.validate_pov调用后根据triage来调用这两个record函数。请检查下在CrashTriggerAgent不运行的情况下，PathExplorationAgent会不会在每次生成test case后除了进行覆盖率插桩的执行，也进行sanitizer插桩的执行（validate_pov），并进行正确的record操作。

% Done.
% \item Unique PoV num (\ourtool <---> baselines) $\checkmark$
% Give up...
% \item LLM interaction num.  
% \item How long are the crash traces of solved tasks?
% \item For the failed tasks, did they reach the vulnerable location?

% \subsection{Case Study}

% TBD

%% file: sections/6_discussion.tex
\section{Discussion}

\subsection{Comparison with Directed Fuzzing} % LLM Agents vs. Fuzzing % Comparison with Fuzzing

Directed greybox fuzzing (DGF) is fundamentally designed to optimize reachability objectives, typically by minimizing a distance metric between runtime states and a target program location, and has proven effective for coverage-oriented vulnerability discovery. However, PoV reproduction is a more goal-directed task that requires not only reaching a vulnerable location, but also satisfying a conjunction of semantic feasibility constraints—often involving non-trivial data dependencies, protocol invariants, or stateful checks—that are not explicitly encoded in the control-flow graph. Importantly, our method does not attempt to replace fuzzing nor to enhance it with more sophisticated heuristics. Instead, it reformulates PoV generation as an interactive hypothesis–verification process: the agent iteratively proposes candidate inputs, observes concrete execution feedback and failures, and refines its hypotheses based on semantically grounded feedback. This process is closer to program debugging than to stochastic exploration, and therefore addresses a complementary problem setting to fuzzing.

% Other discussable points:
% 1. The orders-of-magnitude difference in the number of generated test cases requires specialized designs for LLM-agent solutions and highlights their unique advantages.
% 2. Fuzzing is heavier than LLM-agent solutions to switch among many different history commits.

\subsection{Threats to Validity}

\parabf{Internal Validity.}
% System implementation correctness. % “这是潜在威胁，我们采取了 X/Y/Z 缓解”
% Uncontrolled randomness in the agentic system may introduce variability in outcomes.
% Temperature超参数设置：只有RCA和两个generation，温度不是0.1。
\emph{System implementation correctness} is a potential threat, as \ourtool integrates multiple sub-agents and interacts with a complex execution environment (e.g., coverage queries and Python scripts). However, the correctness of our reported results ultimately depends on whether a generated PoV triggers a sanitizer-confirmed crash. This final validation is performed using a well-tested sanitizer-instrumented binary (\FormulaIn{P_{san}}) and guarded by the lightweight ``Instr Effectiveness Check'' (Figure~\ref{fig:code_instr}). Consequently, implementation bugs in earlier components would at worst lead to PoV generation failures rather than false positives. We further manually inspected most generated PoVs and their execution logs.

\emph{Uncontrolled randomness} may also introduce variability in outcomes. We mitigate this by using a low temperature (0.1) for most LLM interactions, and a higher temperature (0.7) only during test case and PoV generation in Phases III and IV to encourage diversity. However, these stages typically involve $N$ iterative rounds that generate multiple input candidates, which helps amortize the impact of randomness across iterations. Overall, \ourtool's pre-structured workflow and timely intermediate validation reduce the impact of nondeterministic behavior in the agentic system.

% {LLM data leakage and knowledge cutoff.}
% Knowledge cutoff (compare the similarity of we generated PoVs and the ground-turth PoVs) (Maybe add an appendix)

\vspace{5pt}
\parabf{External Validity.}
% External validity concerns the extent to which our findings generalize beyond the evaluated benchmarks and vulnerability settings. 
% \revision{LLM data leakage. Finish this after experiments.}
% Knowledge cutoff (compare the similarity of we generated PoVs and the ground-turth PoVs) (Maybe add an appendix)
%
A potential threat to external validity is \textit{LLM data leakage}, since SEC-bench consists of public CVEs for which PoVs may already exist online, raising the risk that the backbone model reproduces memorized solutions rather than synthesizing new ones. We provide two qualitative indications that this effect is limited. First, under the same backbone model, our agent solves substantially more vulnerabilities than the OpenHands baseline, suggesting gains from the agent design rather than latent recall. Second, as shown in Table~\ref{tab:pov-similarity}, our dual-metric similarity analysis shows low overlap between generated and ground-truth PoVs (average score 0.0296, max 0.1320 across 24 successfully solved CVEs), indicating that our system produces structurally and lexically distinct PoVs rather than reproducing reference implementations.

Another potential threat arises from \emph{benchmark quality}. For example, the sanitizer report for imagemagick.cve-2019-13301 contains multiple types of crashes with several distinct backtraces, which may complicate the final triage and validation process. In such cases, validated PoVs may be confused with variant PoVs that trigger different but related crashes. However, this issue does not affect the core experimental outcomes, as all reported successes are still grounded in sanitizer-confirmed crashes consistent with the benchmark specification.

Regarding supported \emph{vulnerability types}, our evaluation currently focuses primarily on memory vulnerabilities. Nevertheless, \ourtool is not inherently limited to this class. It can be extended to other C/C++ vulnerability types, such as path traversal or command injection, by adapting the crash-triggering prompts and replacing the final PoV validation environment accordingly, following the approach used in FaultLine~\cite{nitin2025faultline}.

\begin{table}[h]
\centering
\small
\begin{threeparttable}
\caption{PoV Similarity Between \ourtool Generated and Ground-Truth PoVs}
\label{tab:pov-similarity}
% \begin{tabular}{lccc}
\begin{tabular*}{0.95\columnwidth}{@{\extracolsep{\fill}} l c c c}
\toprule
\textbf{Statistic} & \textbf{GramSim} & \textbf{ChunkSim} & \textbf{Score} \\
\midrule
Average & 0.0413 & 0.0022 & 0.0296 \\
Max     & 0.1886 & 0.0479 & 0.1320 \\
Min     & 0.0000 & 0.0000 & 0.0000 \\
\bottomrule
\end{tabular*}
\begin{tablenotes}[flushleft]
  \footnotesize
  \item \textit{Note:} Higher values indicate a greater degree of similarity between the generated and ground-truth PoVs.
\end{tablenotes}
\end{threeparttable}
\end{table}

%% file: sections/7_related_work.tex
\section{Related Work}

\parabf{LLM-agent-based Methods.} 
Recent studies have explored the use of LLM-based agentic systems for various PoV generation tasks. 
For vulnerability reproduction in C/C++ programs, PBFuzz~\cite{zeng2025pbfuzz} guides LLMs to generate parameterized input generators and randomly samples the constrained input space to produce PoVs. However, the generated input generators suffer from construct validity bias. They also struggle to craft malformed inputs in certain cases, a task that pure LLM-agent approaches can accomplish with much greater flexibility.
Another recent work~\cite{sapia2026scaling} proposes a similar framework that provides coverage feedback to LLM agents to better address the reachability gap. Nevertheless, its agent focuses solely on reaching target location in function-level, while subsequent crash triggering still relies on existing fuzzers~\cite{galland2025invivo}. In contrast, \ourtool also integrates coverage feedback into LLM reasoning to resolve vulnerability-triggering constraints and supports end-to-end PoV generation, making the two methodologies fundamentally different. 
FaultLine~\cite{nitin2025faultline} does not target memory vulnerabilities; instead, it proposes an agentic workflow to generate PoVs for taint-style vulnerabilities in C/C++ programs (e.g., command injection). However, it merely collects low-level path constraints and directly delegates them to LLMs for solving, which overlooks high-level vulnerability semantics.

Several other agent-based works have also explored the more generalized PoV generation task, but their focus differs from the research scope of this paper.
EnIGMA~\cite{abramovichenigma} builds upon the SWE-agent~\cite{yang2024swe} to solve Capture The Flag (CTF) challenges, enabling interaction with both a debugger and a server connection tool.
PwnGPT~\cite{pwngpt2025} is an LLM-based framework for automatic exploit generation targeting CTF pwn challenges, which partially overlaps with the PoV generation task.
However, synthetic CTF tasks are generally less complex than real-world CVEs, and other categories of CTF challenges (e.g., reverse and crypto) have limited relevance to PoV generation for C/C++ vulnerabilities. 
PoCGen~\cite{simsek2025pocgen} can generate and validate PoC exploits for vulnerabilities in npm packages with the help of LLM, while we focus on different programming language.
There are also research efforts~\cite{zhu2024teams, zhu2025cve, liu2025webvuln} that investigate the reproduction and exploitation of web vulnerabilities, which are substantially different from memory vulnerabilities. 

% \smallskip
\vspace{3pt}
\parabf{Fuzzing-based Methods.} 
Fuzzing-based methods\cite{AFL, AFLplusplus-Woot20}, especially directed greybox fuzzing\cite{bohme2017directed, you2017semfuzz, chen2018hawkeye, lee2021constraint, huang2022beacon, li2024sdfuzz, bao2025alarms, chen2025idfuzz}, have also been widely adopted to address PoV generation tasks.
SDFuzz~\cite{li2024sdfuzz} introduces a target states-driven directed fuzzing approach that derives high-value program states from vulnerability descriptions to guide seed exploration, combined with selective instrumentation to prune irrelevant execution and accelerate reaching targets.
Lyso~\cite{bao2025alarms} leverages correlations among static-analysis alarms and decomposes each target into multiple semantic steps to guide fuzzing toward true positives across multiple related vulnerability sites. 
However, recent studies~\cite{huang2024directed, bao2025alarms, zeng2025pbfuzz} reveal that state-of-the-art undirected general fuzzers (e.g., AFL++) can even outperform directed fuzzers on the given vulnerability benchmarks such as Magma~\cite{hazimeh2020magma}. This is partly because the distance metrics adopted by current fuzzing methods cannot perfectly capture the actual distance to vulnerable states, due to inherent limitations of static program analysis. 
For example, AFLGopher~\cite{bai2025aflgopher} points out that the distance calculation in most existing approaches is feasibility-unaware, which may even mislead the fuzzer with inaccurate feedback.
In contrast, \ourtool mimics the behavior of human vulnerability analysts by reasoning solely based on faithful execution feedback rather than potentially imprecise distance metrics, thereby enabling effective PoV construction.

Several works~\cite{stephens2016driller, alhuzali2016chainsaw, li2025colorgo} combine fuzzing with symbolic execution to satisfy the constraints required to reach specific vulnerable locations. However, inherent issues of symbolic execution (e.g., state explosion) may limit its scalability.
Recent work~\cite{bin2025attention, feng2025randluzz, zeng2025pbfuzz} has also leveraged large language models to enable more semantics-aware directed fuzzing, for example by using LLMs to guide distance metrics or to synthesize reachable seeds and mutators. 
In future work, we plan to explore the integration of \ourtool with traditional AFL-style fuzzing techniques, such as performing further fuzzing based on the test cases generated by our LLM agent.

\vspace{3pt}
\parabf{AIxCC CRSs}
DARPA’s Artificial Intelligence Cyber Challenge (AIxCC)~\cite{AIxCC2023} has attracted significant attention and participation from both academia and industry. 
In the final round, seven teams were required to each submit a cyber reasoning system (CRS) that autonomously performs the end-to-end pipeline of vulnerability discovery, triggering, and patching to earn scores. 
These CRSs also adopt PoV generation techniques similar to those studied in this paper, which consist of two complementary pipelines: enhanced fuzzing and LLM-based generation~\cite{zhang2026sokdarpasaicyber}.
However, directly applying them to our problem setting is non-trivial.
% 1.Fuzzing but not directed. 
First, in the AIxCC competition, vulnerability locations are not provided in advance. In most finalist team CRSs~\cite{kim2025atlantis, fuzzingbrain, trailofbits_blog} that rely on undirected fuzzing techniques such as AFL++, vulnerability discovery and PoV generation are tightly coupled and performed simultaneously by the fuzzing engine. This design makes them unsuitable for specialized PoV generation tasks that require triggering vulnerabilities of specific types at known locations.
% 2.Cost issue
Second, the cost overhead significantly limits the practicality of AIxCC CRSs in real-world scenarios. 
In the final competition, organizers provided each team with \$50,000 in LLM API credits and \$85,000 in Azure compute resources~\cite{kim2025atlantis}. 
Therefore, although Team Theori’s CRS appears to be directly applicable to the vulnerability reproduction task studied in this paper, as it adopts a pipeline of static analysis to locate potential vulnerabilities followed by LLM-based PoV generation~\cite{theori2025aixcc}, it remains difficult to deploy in practice due to cost constraints.

Consequently, the exploration of effective yet cost-efficient solutions still remains an open and worthwhile research direction.

%% file: sections/8_conclusion.tex
\section{Conclusion}
This paper presents \ourtool, an agentic framework that addresses the challenge of automated PoV generation by bridging the gap between static semantic reasoning and dynamic execution.
Unlike prior approaches that rely on blind mutation or open-loop generation, \ourtool formulates the task as an iterative hypothesis--verification--refinement process.
By systematically translating low-level execution states into source-level semantic feedback, our approach enables LLMs to effectively reason over both path reachability and complex vulnerability-triggering conditions.
Our evaluation on the SEC-bench dataset demonstrates that this execution-grounded design substantially outperforms existing LLM-based baselines under fixed budget constraints, solving up to 52.8\% more CVE tasks than the previously best-performing baseline and successfully reproducing many vulnerabilities that were previously unresolved.